\documentclass[twocolumn]{aastex63}
\usepackage[utf8]{inputenc}
\usepackage{amsmath}
\usepackage{graphicx,color}
\usepackage{natbib}
\usepackage[varg]{txfonts}
\usepackage{url}
\usepackage{xspace}
\usepackage{verbatim} 

\newcommand\va{{V808~Aur}\xspace}

\begin{document}

\title{An Eccentric Planet Orbiting the Polar V808~Aurigae}

\author[0009-0008-3152-6773]{McKenna Leichty}
\affiliation{Department of Physics and Astronomy, University of Notre Dame, Notre Dame, IN 46556, USA}

\author[0000-0003-4069-2817]{Peter Garnavich}
\affiliation{Department of Physics and Astronomy, University of Notre Dame, Notre Dame, IN 46556, USA}

\author[0000-0001-7746-5795]{Colin Littlefield}
\affiliation{Department of Physics and Astronomy, University of Notre Dame, Notre Dame, IN 46556, USA}
\affiliation{Bay Area Environmental Research Institute, Moffett Field, CA 94035 USA}

\author[0000-0003-3441-9355]{Axel Schwope}
\affiliation{Leibniz-Institut für Astrophysik Potsdam (AIP), An der Sternwarte 16, 14482 Potsdam, Germany}

\author[0000-0002-8360-092X]{Jan Kurpas}
\affiliation{Leibniz-Institut für Astrophysik Potsdam (AIP), An der Sternwarte 16, 14482 Potsdam, Germany}
\affiliation{Potsdam University, Institute for Physics and Astronomy, Karl-Liebknecht-Straße 24/25, 14476 Potsdam, Germany}

\author[0000-0002-5897-3038]{Paul A. Mason}
\affiliation{New Mexico State University, MSC 3DA, Las Cruces, NM, 88003, USA}
\affiliation{Picture Rocks Observatory, 1025 S. Solano Dr. Suite D., Las Cruces, NM 88001, USA}

\author{Klaus Beuermann}
\affiliation{Institut für Astrophysik, Georg-August-Universität, Friedrich-Hund-Platz 1, 37077 Göttingen, Germany}

\begin{abstract}

We analyze 15 years of eclipse timings of the polar V808~Aur. The rapid ingress/egress of the white dwarf and bright accretion region provide timings as precise as a few tenths of a second for rapid cadence photometric data. We find that between 2015 and 2018, the eclipse timings deviated from a linear ephemeris by more than 30~s. The rapid timing change is consistent with the periastron passage of a planet in an eccentric orbit about the polar. The best fit orbital period is 11$\pm 1$ yr and we estimate a projected mass of $Msin(i)=6.8\pm 0.7$ Jupiter masses. We also show that the eclipse timings are correlated with the brightness of the polar with a slope of 1.1~s/mag. This is likely due to the change in the geometry of the accretion curtains as a function of the mass transfer rate in the polar. While an eccentric planet offers an excellent explanation to the available eclipse data for V808~Aur, proposed planetary systems in other eclipsing polars have often struggled to accurately predict future eclipse timings.
 
\end{abstract}

\keywords{cataclysmic variables---AM Herculis stars---exoplanet detection methods---eclipsing minima timing methods---stellar magnetic fields}

\section{Introduction}

The precise timing of short-period eclipsing binaries provides a sensitive method to search for planets or substellar companions orbiting the system \citep[e.g.][]{beuermann10}. Over an orbit, a planet shifts the center of mass of the close pair resulting in variations in the eclipse times due to the light-travel time effect. The very small physical size of white dwarfs (WD) means that an eclipse in a cataclysmic variable (CV) or pre-CV systems can be measured with excellent precision. Polars (AM~Her stars) are CVs where a magnetic field from the WD disrupts the formation of an accretion disk. Gas donated from the secondary is funneled to the vicinity of the magnetic poles of the WD, generating a very bright and compact emission region. Thus, eclipsing polars are excellent systems to search for planets or additional stellar components.

Caution is required in interpreting changes in eclipse timings in magnetic CVs. Eclipse timings from the polar HU~Aqr initially suggested multiple planet candidates orbiting the close binary \citep{qian11}. However, extending the observation baseline revealed timing variations that could not be explained solely by a multiple planet hypothesis \citep{schwope_thinius14,schwope18}. The source of this timing variability remains uncertain. 

Here, we investigate the eclipse timings of the polar V808~Aur \citep[aka CSS081231:071126+440405;][]{thorne10}. V808~Aur is a polar with observational properties that are similar to HU~Aqr. Both systems show deep primary eclipses and sometimes pre-eclipse ``dips" caused by the self-eclipse of a magnetic accretion curtain. Both polars have orbital periods very close to two hours. HU~Aqr and V808~Aur display large variations in their mass transfer rate on time scales of weeks to months \citep{schwope15}. 

A study of eclipse timings in V808~Aur by \citet{schwope15} did not detect significant variations in data obtained between 2008 and 2015. Here, we analyze a new set of very precise eclipse times obtained between 2008 to 2023.

\begin{figure}
    \centering
    \includegraphics[width=\columnwidth]{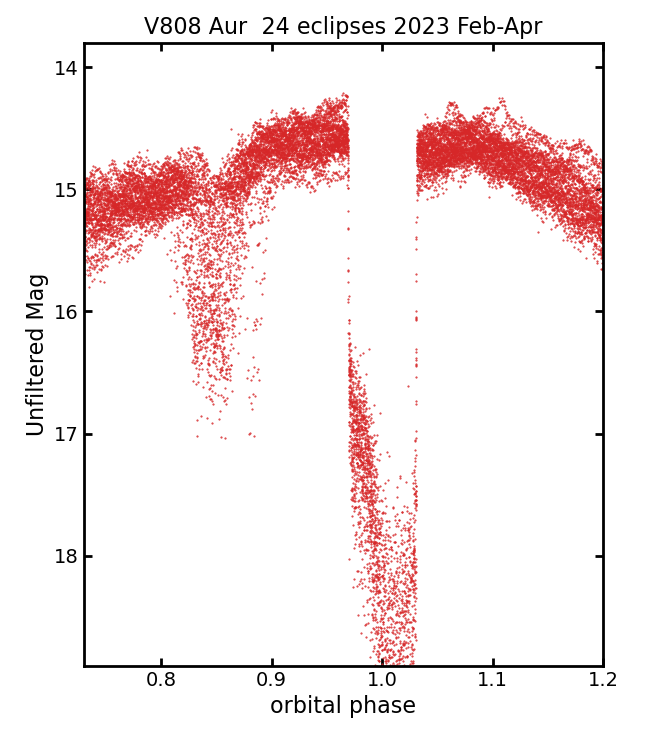}
    \caption{Light curves of V808~Aur obtained with the SLKT in early 2023. The photometric time series were taken with 3.0~s exposures and a 3.6~s cadence. The data are phased on the binary orbital period and 24 eclipses over 13 nights are shown. The brightness of the system was relatively stable over the three months of observations. Pre-eclipse dips are visible between phases 0.8 and 0.9. An accretion curtain remains partly visible during the WD eclipse and can be seen fading in brightness within the main eclipse.}
    \label{newkriz}
\end{figure}

\section{Data}

The times of photometric measurements are referenced to the center of the exposure and converted to BJD (Barycentric Julian Day in the Barycentric Dynamical Time (TDB) system). In our analysis, we include 11 eclipse measurements from \citet{schwope15} that had estimated timing errors of less than 3s. The full set of eclipse timings are given in Table~\ref{timings}.

\subsection{TESS}

TESS observed V808 Aur during sector 47 (s47) with a 120~s cadence starting Dec 31, 2021. The 2~min cadence is not ideal for individual eclipse timing estimates as the 7-minute-long eclipses are poorly sampled. Therefore, we estimated an average time of mid-eclipse by phasing the TESS photometry to the \citet{schwope15} ephemeris and determining the offset of the mid-eclipse from the predicted time. We then converted the measured phase offset to the time of mid-eclipse for a representative eclipse located near the middle of the TESS sector. 

\subsection{Krizmanich Telescope Observations}
\label{sec:kriz}
The Sarah L. Krizmanich Telescope (SLKT) is located on the University of Notre Dame campus and features a 0.8-m (32-in diameter) primary mirror. In January of 2022, a new CMOS detector replaced the original SBIG CCD. We used the CMOS detector to observe V808~Aur from November 2022 to May 2023 using exposure times ranging from 3 to 8~s. These were unfiltered, and each time series usually spanned 2-3~hr with many capturing multiple eclipses.

SLKT observations of V808~Aur from 2017 to 2021 used the SBIG CCD. We corrected the times recorded in the headers to the center of the exposure and added a shutter lag correction.

\subsection{AIP observations}

\va was observed with the 70-cm reflector located at the Babelsberg site of the Leibniz-Institut f\"ur Astrophysik Potsdam (AIP). The telescope is equipped with a new back-illuminated sCMOS camera which allows fast readout at rather low noise. Observations were performed in white light on two nights on February 8 and 28, 2023. The weather conditions were good. On February 8, 2309 single exposures were taken with an integration time of 3~s, resulting in a time resolution of 3.7~s. During this night, three eclipses were covered.  
A further two eclipses were observed on February 28, when 7863 exposures with integration times as short as 1~s were obtained (time resolution 1.7~s).  Differential photometry on dark-subtracted and flatfield corrected images was performed to determine the relative magnitudes of \va. The comparison star used is Gaia EDR3 950114080899045248 at RA=07:11:35.505, and DE=44:04:03.27. It has a mean magnitude $G=13.901$ and $BP-RP = 1.166$.

\subsection{MONET/North, Calar Alto, and Las Cumbres Observatory}

A total of 60 eclipses of V808~Aur were observed in white light
between 2009 and 2014 with the 1.2-m MONET/N telescope located at the
McDonald Observatory/Texas. An additional nine eclipses were acquired
with the 80-cm Schmidt telescope at the Calar Alto Observatory/Spain,
the 2-m Faulkes/N, the 1-m telescopes of the Las Cumbres
Observatories at the sites on Hawaii, Teide/Teneriffa, and
McDonald Observatory (this set of data referred to as ``MONET/N+''). All measurements employed a 10~s cadence with a 7~s
readout time. The data were corrected for dark current and flat-fielded
in the usual way. Relative photometry was performed on all frames with
respect to the same non-variable comparison star. The observed light
curves were modeled by the eclipse of a uniformly emitting circular
disk \citep{backhaus12}. The mid-eclipse times were
estimated to have a standard deviation
of $\sigma\!=\!2.6$~s. Therefore, a systematic error of 2.0~s was
added in quadrature to the formal fit errors of the mid-eclipse times.


\subsection{McDonald Observatory}
V808 Aur was observed with the McDonald Observatory 2.1-m telescope on 20 nights from February 2018 to January 2023. A broad-band (BVR) filter was used with 1~s to 5~s exposures, depending on seeing quality, with no dead time. All of the McDonald observations were obtained during the high brightness state. A total of 33 eclipses were acquired. The mid-eclipse times were measured using the mid-points of linear fits to both the ingress and egress. Other methods were found to give similar results. These eclipses are labeled ``McDonald'' in the $O-C$ diagram of Figure \ref{system}.
\label{sec:McDonald}

\begin{figure}
    \centering
    \includegraphics[width=\columnwidth]{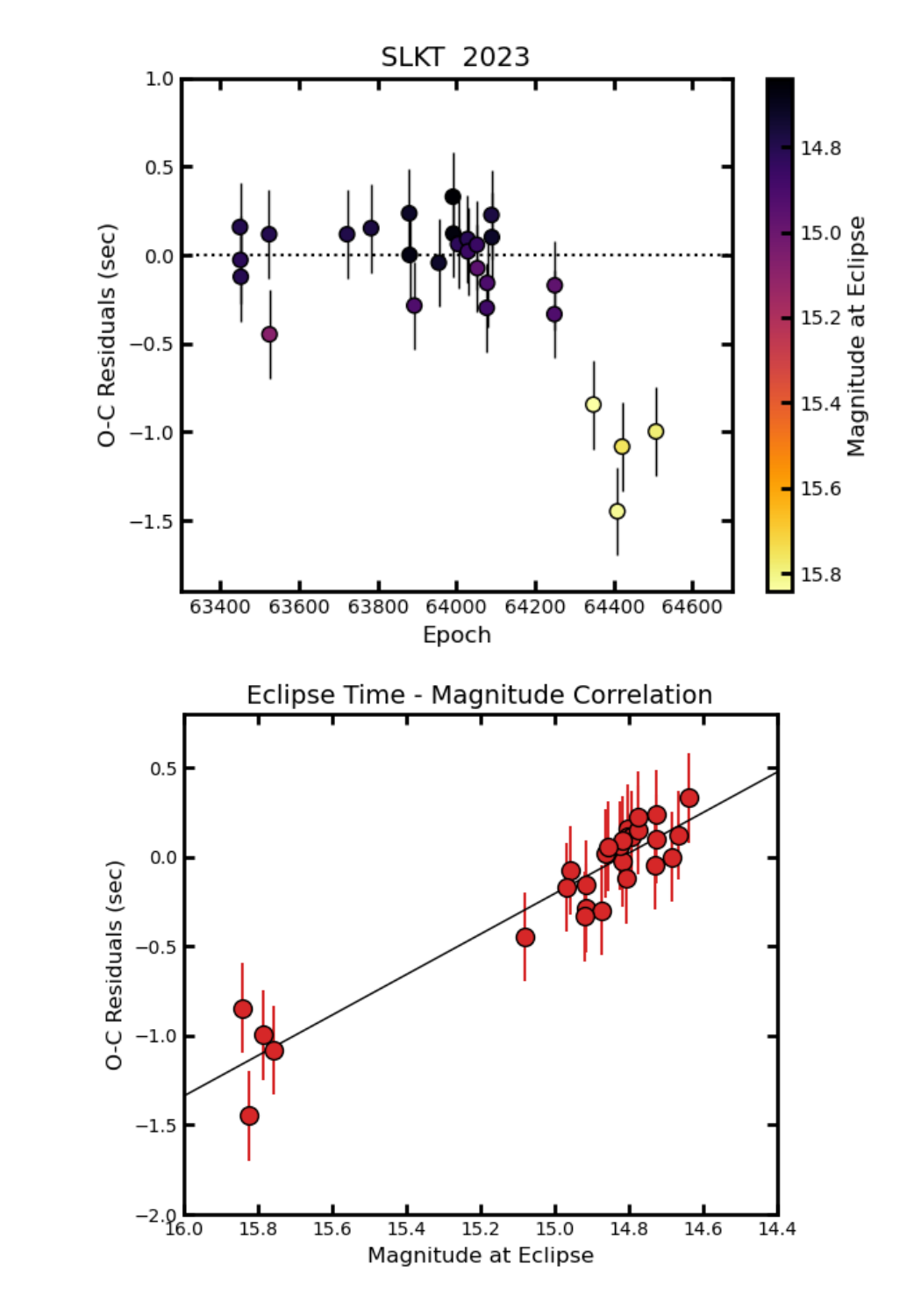}
    \caption{{\bf Top:} The timing residuals for the eclipses observed by SLKT in 2023 using a 3~s exposure time. The variance of the timing, when the system was bright, corresponds to $\sigma =0.21$~s. Starting in 2023 May, the star faded in brightness, and times of mid-eclipse were measured to come 1~s earlier. {\bf Bottom:} Eclipse time residuals plotted as a function of the average magnitude before and after the eclipse. Even small variations in brightness appear well correlated with the timing residuals. The solid line shows a linear fit with a slope of 1.1~s/mag.   }
    \label{slkt}
\end{figure}

\section{Analysis}

Times of mid-eclipse were estimated using the same methods as \citep{schwope15}. Ingress and egress times were referenced to the moment when the flux reached half the brightness at the start/end of the eclipse and the time of mid-eclipse was calculated by averaging the two measurements. When V808~Aur is in a bright state, an accretion curtain\footnote{Also called the magnetically guided portion of the accretion stream in \citet{schwope15}. } remains visible during the first half of the primary eclipse (see Figure 11 in \citet{schwope15}). This is seen in the light curve shown in Figure~\ref{newkriz} as a slow fading within the eclipse from $\approx 16.8$~mag to fainter than 18~mag. That is, the sharp ingress represents the WD and accretion spot moving behind the secondary star, while the curtain arcs out of the orbital plane and takes several minutes to be fully eclipsed. The roughly two-magnitude drop in brightness at ingress implies that the curtain contributes about 15\%\ of the flux at the onset of the eclipse. Variations in the curtain geometry and brightness correlated with the mass transfer rate may impact the eclipse timings, although this is expected to be a small effect.  

\subsection{Timing Precision}

Early in 2023, we obtained 28 eclipse timings using 3.0~s exposures (3.6~s cadence) with the SLKT. In the first 24 eclipses, the brightness of V808~Aur, estimated by averaging the magnitudes just before and after an eclipse, varied by less than one magnitude (see Figure~\ref{newkriz}). This provided a consistent set of timings that we use to estimate the precision of the eclipse timings. Using the ephemeris from \citet{schwope15}, we calculated the difference between the observed time of mid-eclipse, and the ephemeris prediction, $O-C$. We calculated the least-squares linear fit to the $O-C$ points for the brightest 24 eclipses and subtracted the fit to obtain the residuals. The $O-C$ residuals are plotted in the top panel of Figure~\ref{slkt} and the timings of the faintest four points are clearly offset by approximately one second from the brightest points. The residuals of the brightest 24 measurements correspond to a standard deviation of $\sigma =0.20$~s. This approximates the random measurement uncertainty for the fastest cadence data.

As seen in Figure~\ref{newkriz}, the eclipse ingress shows a rapid fading of 2~mag in less than 4~s, suggesting that the bright source being eclipsed is unresolved with the 3.6~s cadence. This is also true of the eclipse egress where the final 2~mag rise takes less than 4~s. We conclude that the light from the main accretion pole dominates the shape of the eclipse.

\subsection{Timing Variations Correlated with Brightness}

The final four eclipse observations of the 2023 season caught the V808~Aur system fading by about a magnitude from a few days earlier. The SLKT eclipse timings also show a shift of about 1~s after the system brightness faded. A change in eclipse timing with varying brightness is expected since the position of the accretion curtain shifts with mass transfer rate \citet{schwope15}. The abundant SLKT eclipse timings capture the timing changes with brightness as shown in the lower panel of Figure~\ref{slkt}. In fact, there is a tight correlation between eclipse residuals and system brightness with a slope of 1.1~s/mag. The standard deviation about the linear fit is only $\sigma = 0.14$~s, meaning that the 0.20~s measurement uncertainty estimated in the previous section contained a significant contribution from timing variations correlated with the system brightness.  Historically, the brightness of the system around eclipse ranges over 3~mag, although the slope of the correlation is only estimated here for a one-magnitude variation at a high mass transfer rate. The impact of changes in brightness on the eclipse times is relatively small, and here, we take it into account by adding a systematic uncertainty of 1.0~s in quadrature to the other errors.

\subsection{Systematic Timing Uncertainties}

\begin{figure}
    \centering
    \includegraphics[width=\columnwidth]{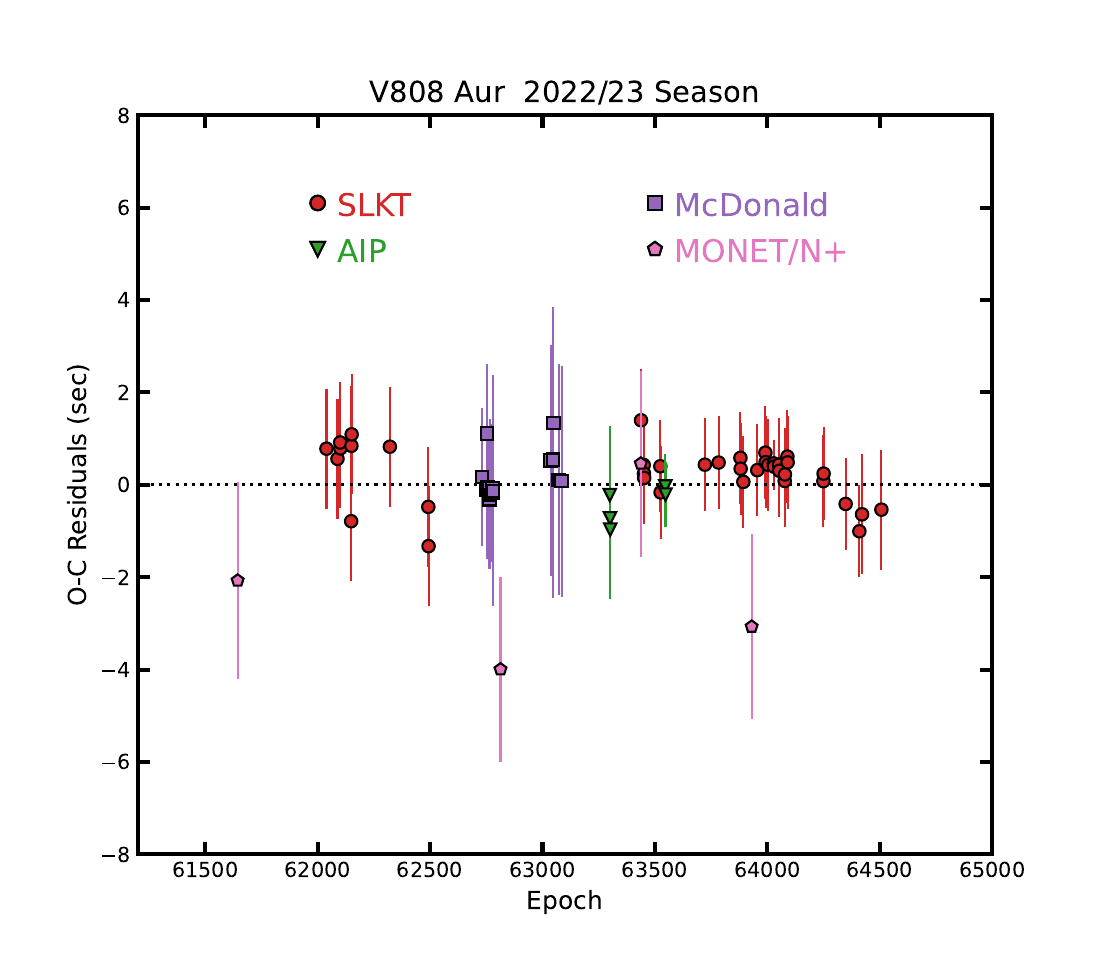}
    \caption{Timing residuals from the 2022/23 observing season from four telescopes. A weighted linear fit has been subtracted to remove any trend from the assumed orbital period. The scatter represents a combination of statistical uncertainties and systematic errors resulting from clock errors and accretion rate variations in the polar. The variance of the data is consistent with a $\sigma = 0.90$~s.  }
    \label{system}
\end{figure}

Systematic errors in the eclipse timings may come from clock errors, uncorrected shutter lag, and possibly other unanticipated sources. To estimate the systematic uncertainties, we compare the timings from the four telescopes described in Sections~\ref{sec:kriz} to \ref{sec:McDonald} during the 2022/23 observing season. Using the \citet{schwope15} ephemeris, we calculated the $O-C$ times from the four systems and then removed any trends by subtracting a weighted linear fit. The results are shown in Figure~\ref{system}. The error bars were estimated from the eclipse fitting algorithm. 

The standard deviation of the $O-C$ residuals is $\sigma =0.90$~s. The $\chi^2$ parameter for these data is 20.1 with 61 degrees of freedom. Thus, the assigned errors are likely over-estimated. Because we have less control over the older photometry going back 15 years, we will adopt the assigned uncertainties.

\subsection{Long-Term Eclipse Timing Variations}

Figure~\ref{o-c} shows 207 observed times of mid-eclipse, compared with the ephemeris derived by \citet{schwope15}. It is clear that between 2014 and 2016, there was a rapid deviation from the predicted timing resulting in a jump of over 50~s in the $O-C$ values. The jump is suggestive of a light-time shift caused by the periastron passage of a third body in a highly eccentric orbit. Alternatively, an unknown process acting within the polar binary might generate large $O-C$ variations as is seen in HU~Aqr  \citep[e.g.][]{schwope18}.

\section{Discussion}

\subsection{Polynomial Fits to the Timing Measurements}

After 2015, a linear ephemeris does not appear to be a good representation of the V808~Aur eclipse times. Because there are two changes in the derivative seen in Figure~\ref{o-c}, a 3rd-degree polynomial (cubic) is the minimum number of terms that can represent the timing changes. However, minimizing the $\chi^2$ parameter using a cubic polynomial results in a poor fit with $\chi^2>4000$ for 203 degrees of freedom. Adding terms to the polynomial model does improve the $\chi^2$ parameter, but the resulting fits remain poor. The rapid change in the eclipse timings between 2015 and 2017 is not easily modeled using a polynomial.

To estimate the rate that the orbital period is changing, $\dot{P}$, we divide the $O-C$ curve into two sections based on the sign of its curvature. We fit a quadratic function to eclipses measured earlier than 2017 and find a $\dot{P}=+2.48 \pm 0.05 \times 10^{-11}$. For eclipses after 2014, the quadratic fit gives a period derivative of $\dot{P}=-2.28 \pm 0.08 \times 10^{-11}$. The respective $\chi^2$ values for these fits are 2047/85 dof and 2492/168 dof. 

\subsection{Orbit Elements of a Third Object}

\begin{figure}
    \centering
    \includegraphics[width=\columnwidth]{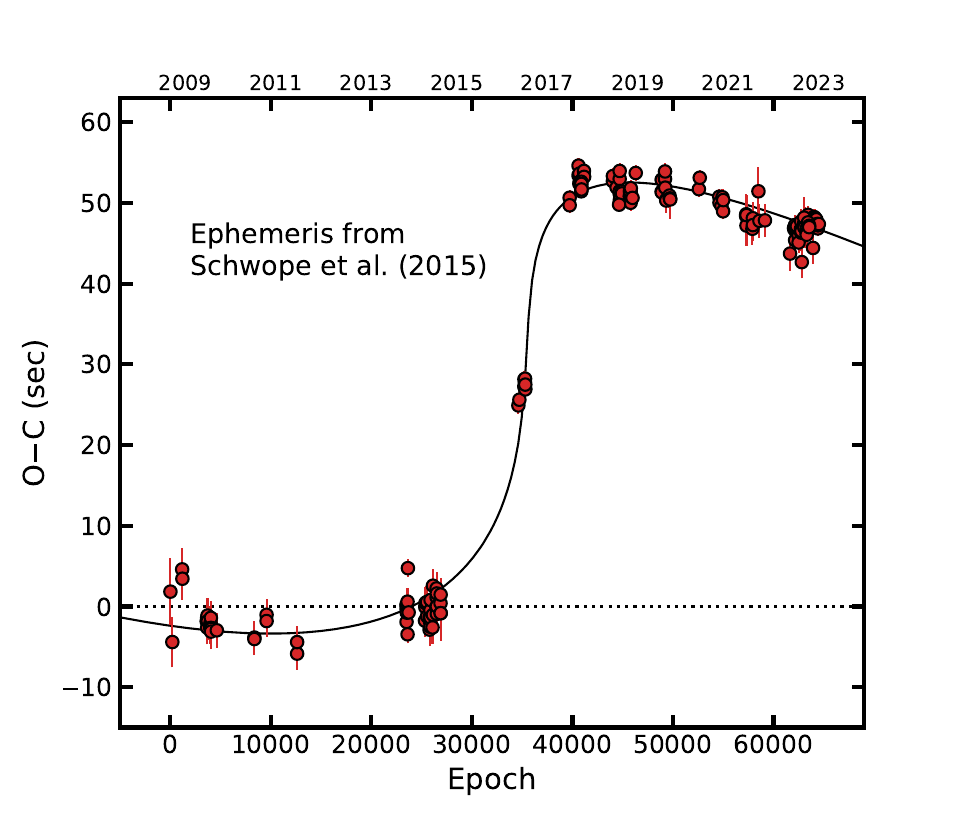}
    \caption{The $O-C$ times calculated using the \citet{schwope15} ephemeris. The solid line is the best-fit orbit of a third body where the binary period is fixed to the \citet{schwope15} period. The orbital period of the third body is 38~yr.  }
    \label{o-c}
\end{figure}

Assuming that the observed $O-C$ measurements result from a third body orbiting the polar, there appears to have been only one periastron passage over the past 14 years. The variation shown in Figure~\ref{o-c} implies a fairly long orbital period of several decades. However, the shape of the $O-C$ curve is very dependent on the assumed inner binary orbital period, thus neither the binary period, $P$, nor the period of a third body, $P_3$, are constrained without detailed orbital modeling.

There are approximately 60000 eclipses between 2009 and 2023, so millisecond adjustments to the binary orbital period yield $O-C$ variations of the same order as seen in Figure~\ref{o-c}. By rounding the period measured by \citet{schwope15} to the nearest 10~ms, we define a base binary period of $P_0=7030.95$~s. We make millisecond adjustments to the binary orbital period using an offset parameter $\Delta P_0$, so that $P=P_0\; +\; \Delta P_0$. In this convention, the precise binary orbital period estimated by \citet{schwope15} corresponds to $\Delta P_0=6.3321$~ms.

To compare the observed eclipse timings with a third body orbital model, we subtract a linear ephemeris from the observed times of mid-eclipse, $T_C(BJD)$, such that 
\begin{equation}
(O-C)_{data} = T_C -(T_0 + E\; (P_0+\Delta P_0)) \; ,
\end{equation}
where $T_0=BJD\;2454833.207868$ was selected to be the initial epoch from \citet{schwope15}. The binary orbital period, $\Delta P_0$, is a free parameter.

\citet{irwin52} described a method of estimating the orbital parameters of a third body based on light-travel time variations in the eclipse timings of a binary. 

Following the study of V606~Cen by \citet{li2022}, we calculate a model for the eclipse time variations, $(O-C)_{model}$ (their equation 2) but here we do not include a binary period derivative term. The model is given by
\begin{equation}
(O-C)_{model} = T_0 +\Delta T_0 + E\; (P_0+\Delta P_0) + \tau \; ,
\end{equation}
where $\Delta T_0$ adjusts the time of the initial epoch.
The function $\tau$ is the light-time effect term defined by
\begin{multline}
\tau (K,e,\omega,\nu(E_0)) = K\frac{1}{\sqrt{(1 - e^2\cos^2(\omega))}} \\
\times \left[ \frac{1 - e^2}{1 + e\cos(\nu)}\sin(\nu + \omega) + e\sin(\omega) \right],
\end{multline}
where $\tau$ depends on $K$, the semi-amplitude of the light-time effect; the eccentricity, $e$; the longitude of the periastron, $\omega$; and the true anomaly, $\nu$. The true anomaly is calculated from the mean anomaly by summing a series of Bessel functions. The true anomaly is a function of time, and therefore of the binary orbital epoch. Finally, the epoch of periastron passage, $E_0$, sets the phase of $\nu$.

To constrain the orbital parameters, we construct and minimize a $\chi^2$ function,
\begin{multline}
\chi^2(\Delta P_0,\Delta T_0, P_3, K, e, \omega, E_0) = \\ \large\sum_i [(O-C)_{data,i}-(O-C)_{model}]^2/\sigma_i^2 \; ,
\end{multline}
where $\sigma^2$ is the variance of the mid-eclipse time measurement.

As a starting point, we set the binary orbital period to $P_0 + \Delta P_0$ with $P_0=7030.95$~s and $\Delta P_0=6.3321$~ms, the period estimated by \citet{schwope15}, and minimized $\chi^2$ with six free parameters. The results are shown in Figure~\ref{o-c}, and the model matches the data fairly well. The $\chi^2$ parameter from this period is 275 for 201 degrees of freedom (dof). The orbital period of the third body in this fit is 38.4~yr.

\begin{figure}
    \centering
    \includegraphics[width=\columnwidth]{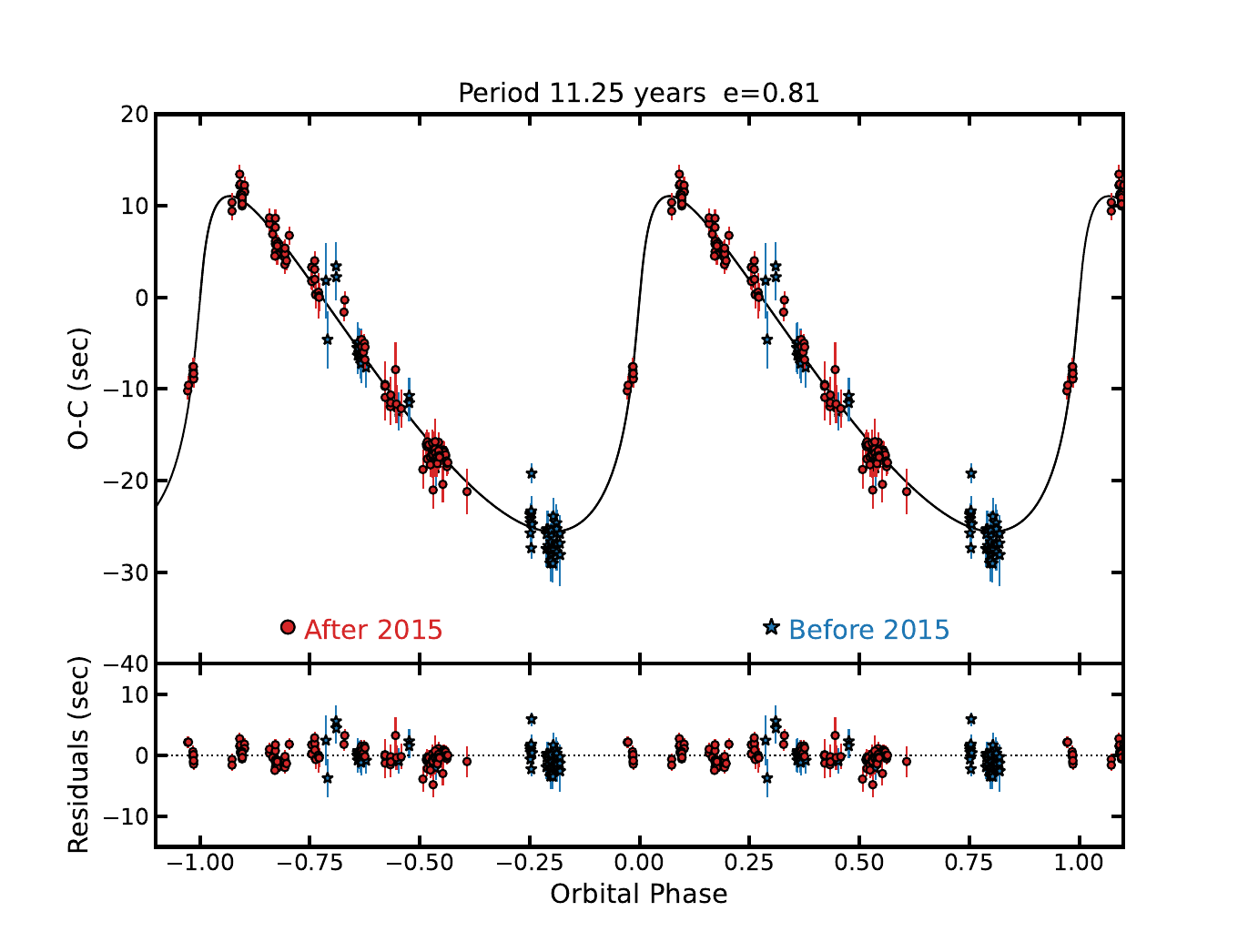}
    \caption{The $O-C$ times calculated from the third body orbit that minimize the $\chi^2$ parameter. The orbital cycle is repeated for clarity. Residuals to the best-fit model are shown in the lower panel. Timings obtained before 2015 are displayed as blue stars, while data taken after 2015 are shown as red circles.  }
    \label{best}
\end{figure}

Increasing the assumed binary orbital period consistently yielded shorter third-body orbits and improved $\chi^2$ parameters. Adding the binary period, $\Delta P_0$, as a free parameter of the fit, the minimum $\chi^2$ was found for a binary orbital period of $\Delta P_0=7.347$~ms ($P=7030.957347$~s). The model parameters that give the minimum $\chi^2=199.9$ for 200 dof are provided in Table~\ref{parameters} and the resulting eclipse timing variations are shown in Figure~\ref{best}. The third body orbit is a rather short 11.25~yr, meaning that the next periastron passage is expected to occur in 2028. Rapid changes in the eclipse timings are likely to be detectable a few years before the periastron passage.

\begin{deluxetable}{lccc}
\centering
\tablecaption{Parameters of the V808 Aur System \label{parameters}}
\tablehead{
\colhead{Parameter} & \colhead{Value} & \colhead{Uncertainty} & \colhead{Units} }
\startdata
Binary Orbital Period$^a$, $\Delta P_0$ & 7.346 & 0.031 & ms \\
Initial Epoch Shift, $\Delta T_0$ & $-7.26$  &  0.32  & s \\
Orbital Period, $P_3$ & 11.25 & 0.99 & yr \\
Semi-amplitude, $K$  & 18.32  & 0.51 & s \\
Eccentricity, $e$  &  0.805  &  0.020  &  { }  \\
Epoch of Periastron, $E_0$  & 36039 & 135 & epochs \\ 
Longitude Periastron, $\omega$  & 14.3 & 0.7 & deg \\
Semi-Major Axis, $a\; sin(i)$ & 0.059 & 0.003 & AU \\
Mass Function, $f(m)$  &  3.31$\times 10^{-7}$  &  $0.69\times 10^{-7}$  &  M$_{\odot}$ \\
Projected Mass$^b$, $M_3\;sin(i)$  & 6.8  &  0.7  &  M$_{jup}$ \\
$\chi^2$/dof   & 199.9/200  &  { }  &   { }
\enddata
\tablenotetext{a}{The binary orbital period, $P=P_0+\Delta P_0$ where $P_0=7030.95$~s}
\tablenotetext{b}{Assuming $M_{tot} = M_1 +M_2 +M_3= 0.9\pm 0.1~{\rm M}_\odot$.}
\end{deluxetable}

\subsection{Properties of the Third Object}

Given the best-fit orbital parameters and the light-time orbit amplitude $K$, we can estimate the projected semi-major axis of the orbit, $a\;sin(i)$ \citep{irwin52},
\begin{equation}
a\; sin(i) = \frac{c\; K}{\sqrt{(1 - e^2\cos^2(\omega))}} \; ,
\end{equation}
where $c$ is the speed of light.

The mass function, $f(M_3)$, for an orbit with significant eccentricity is \citep{tauris06},
\begin{equation}
f(M_3) = {4\pi^2\; \frac{(a\;sin(i))^3}{G\, P_3^2}} (1-e^2)^{\frac{3}{2}} \; ,
\end{equation}
and, $f(M_3)={{M_3^3 sin^3(i)}/{(M_1+M_2+M_3)^2}}$. Approximating the total mass for the system constrains the mass of the third object. From \citet{knigge11},  the average mass of a WD in a CV is 0.79~M$_\odot$ and secondary stars in systems with 2~hr periods are expected to have a mass of 0.15$\pm 0.05$~M$_\odot$. Evaluating equation~6 yields a very small value of order $10^{-7}$~M$_\odot$, meaning that the mass of the third body adds little to the system total. Thus, we will assume the total mass is $M_{tot}=0.9\pm 0.1$. 

\begin{figure}
    \centering
    \includegraphics[width=\columnwidth]{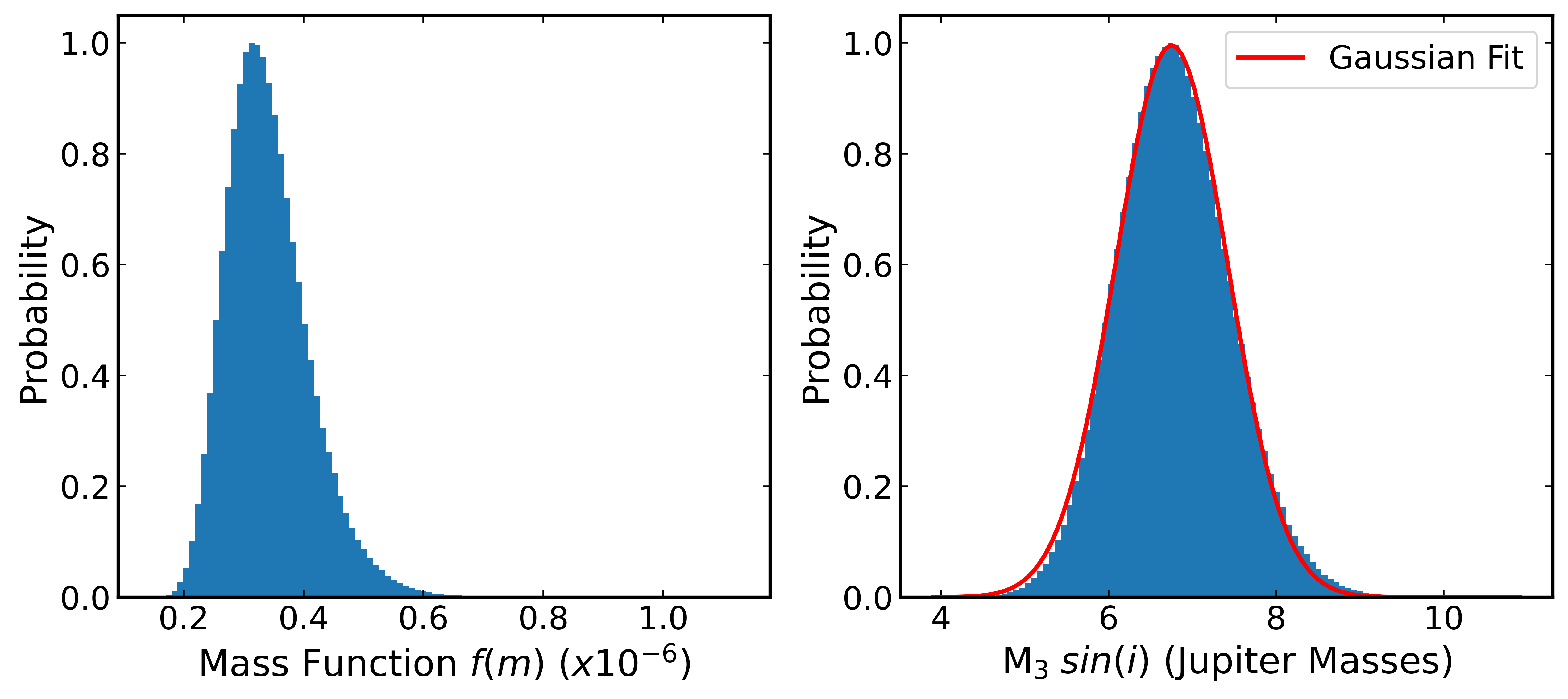}
    \caption{Probability distributions of the mass function and projected mass of the third body orbiting V808~Aur. Fitting a Gaussian to the projected mass distribution produces a mean of 6.8 and $\sigma =0.7$ Jupiter masses.}
    \label{mass_hist}
\end{figure}

Solving for $M_3\;sin(i)$ yields a mass of 0.0065~M$_\odot$, or approximately 7 Jupiter masses for a highly inclined orbit. The probability distributions of the mass function and mass of the third body are shown in Figure~\ref{mass_hist}.

\subsection{Possible Sources of Timing Changes}

While the eclipse timing variations in V808~Aur are well modeled by the passage of a planet-sized body, there may be other explanations for the observed $O-C$ changes. Eclipse timings of the polar HU~Aqr have shown complex changes that are difficult to model with orbits of even multiple planets \citep{schwope18}. There have been attempts to explain orbital variations seen in eclipsing binaries through the Applegate mechanism \citep{applegate92,lanza06}, where a magnetically active late-type companion redistributes its angular momentum via changes in its subsurface magnetic field. Estimates of the connection between changes in the quadrupole moment of the secondary and the resulting variation in the orbital period of the binary suggest that the Applegate mechanism is near the limit of explaining the amplitude of the $O-C$ variations in HU~Aqr \citep{schwope18}. However, a combination of planets and the Applegate mechanism could work for HU~Aqr \citep{bours14}.

The amplitude of the $O-C$ variations in V808~Aur is significantly less than those observed in HU~Aqr, making an origin through the Applegate mechanism more plausible. However, the extremely rapid jump in eclipse timings between 2015 and 2017 implies a large $\dot P\sim 5\times 10^{-11}$ that is difficult for the Applegate mechanism to explain \citep{bours14}. Although, \cite{lanza20} proposed a modification to the Applegate mechanism that might provide a sufficient amplitude to account for these measurements.

\citet{bours14} also considered the possibility that changes in the geometry of the magnetic accretion could generate timing variations as large as 15~s in HU~Aqr. \citet{schwope01} found a relation between the X-ray count rate and phase of the pre-eclipse dips in HU~Aqr, suggesting that the longitude of the curtain and accretion spot depends on the mass transfer rate. In the optical, no pre-eclipse dip is seen when HU~Aqr is faint. Changes in the visibility of the pre-eclipse dip with the brightness of V808~Aur are also observed, implying that its accretion curtain may also shift to low longitudes when the mass transfer rate declines \citep{schwope15}.

Our observation of the correlation between the brightness of V808~Aur and the shift in eclipse timing implies that the changes in accretion geometry could impact the eclipse timings by several seconds, but is unlikely to explain the large jump in $O-C$ observed around 2016. Photometric monitoring (e.g. ATLAS) of V808~Aur during the jump in $O-C$ timing does not show unusual activity. An extended low-state in 2020, and a long high-state in 2022 did not appear to have significantly impacted the eclipse timings over those periods. We conclude that changes in accretion geometry have measurable but minor effects on the eclipse times.

\subsection{Planet Candidates in Other Polars}

Large-amplitude eclipse-timing variations have been observed in at least three other eclipsing polars: HU Aqr \citep{qian11}, DP Leo \citep{qian_dp_leo, beuermann11}, and UZ For \citep{potter11}. For each of these systems, a circumbinary planetary system has been proposed as one possible explanation for the peculiar eclipse timings, and a commonality has been that these planetary models struggle to predict with high accuracy future times of mid-eclipse (HU Aqr: \citealt{schwope_thinius14,schwope18}; DP Leo: \citealt{boyd}, and UZ For: \citealt{khangale}). This suggests that a combination of factors, including a planetary system, the Applegate mechanism, and/or an unknown mechanism of angular momentum loss, collectively influence the eclipse timings \citep{schwope18}. \citet{pulley22} reached a similar conclusion in a study of seven post-common-envelope binaries that have been proposed to host circumbinary planets. They found that the hypothesized planetary systems failed to accurately follow the predicted eclipse ephemerides after further observations were made.

The record for confirming circumbinary planets orbiting polars is poor. However, the eclipse timings for V808~Aur display an excellent fit to a highly eccentric third-body orbit that may stand up to the test of time. 

\bigskip
\section{Conclusions}

Using fast cadence photometry, we have measured the eclipse timings of V808~Aur with precision as small as $\pm 0.2$~s. From these precise times, we found a correlation between the brightness of V808~Aur and eclipse timing residuals amounting to a shift of 1.1~s per magnitude. We attribute this correlation to the changes in the accretion geometry and luminosity with variations in the mass transfer rate. 

Based on the \citet{schwope15} ephemeris, we have measured a greater than 50~s shift in the eclipse timings of V808~Aur between 2015 and 2017. The shift is well-modeled by the light-travel time variation caused by a third body in a highly eccentric orbit. The projected mass of the third body is $M_3\; sin(i)=6.8\pm 0.7$~M$_{jup}$, consistent with a fairly large planet. Assuming an average inclination for random distribution of possible orbits, $M_3=8.1$~M$_{jup}$, which is below the typical mass of brown dwarfs \citep{planetlimit}. The best fit to the planet orbit is found for a period of 11~yr. Thus, the next periastron passage is expected in 2028. 

While a planet in a highly eccentric orbit currently provides an excellent fit to the full set of V808~Aur eclipse timings, the binary orbit may be impacted by the Applegate mechanism or related processes. Further observations are needed to confirm the presence of this planet.


\facilities{TESS, SLKT, AIP, MONET/N, Calar Alto, LCO, McDonald}

\bigskip
\begin{acknowledgements}

We thank Paul Breitenstein and Erwin Schwab for contributing the
observations taken at the LCO Observatories and the Calar Alto
Observatory, respectively. P.A.M. acknowledges support from Picture Rocks Observatory and thanks the McDonald Observatory of the University of Texas at Austin, especially John Kuehne for his improvements to the 2.1-m telescope.  We thank the Krizmanich family for support of the Sarah L. Krizmanich Telescope on the University of Notre Dame campus.

Some of the data presented in this paper were obtained from the Mikulski Archive for Space Telescopes (MAST) at the Space Telescope Science Institute. The specific observations analyzed can be accessed via \dataset[DOI]{http://dx.doi.org/10.17909/f9qc-5493}.

\end{acknowledgements}

\newpage
    

\begin{deluxetable}{lccccr}
\centering
\tablecaption{Eclipse Timing Measurements of the V808~Aur System
\label{timings}}
\tablehead{
\colhead{Epoch$^a$} & \colhead{Mid-Eclipse ($T_C$)} & \colhead{Error} & \colhead{Exposure Time} & \colhead{Date of $T_C$} & \colhead{Telescope$^b$} \\ \colhead{ } & \colhead{BJD} & \colhead{s} & \colhead{s} & \colhead{UT} & \colhead{ }}
\startdata
16 & 2454834.5099183 & 4.1 & 10.0 & 2009-01-3.009 & 1 \\
204 & 2454849.8086862 & 3.1 & 10.0 & 2009-01-18.308 & 3 \\
1182 & 2454929.39531 & 2.6 & 1.0 & 2009-04-7.894 & 7 \\
1206 & 2454931.34834 & 2.6 & 1.0 & 2009-04-9.847 & 7 \\
3622 & 2455127.9546506 & 2.7 & 10.0 & 2009-10-23.453 & 3 \\
3670 & 2455131.8607283 & 2.1 & 10.0 & 2009-10-27.360 & 3 \\
3671 & 2455131.9421195 & 2.1 & 10.0 & 2009-10-27.441 & 3 \\
3707 & 2455134.8716875 & 2.2 & 10.0 & 2009-10-30.371 & 3 \\
3721 & 2455136.0109548 & 2.3 & 10.0 & 2009-10-31.510 & 3 \\
3979 & 2455157.0061638 & 2.1 & 10.0 & 2009-11-21.505 & 3 \\
3990 & 2455157.9013081 & 2.0 & 10.0 & 2009-11-22.400 & 3 \\
4002 & 2455158.8778355 & 2.1 & 10.0 & 2009-11-23.376 & 3 \\
4003 & 2455158.9592103 & 2.0 & 10.0 & 2009-11-23.458 & 3 \\
4014 & 2455159.8543536 & 2.4 & 10.0 & 2009-11-24.353 & 3 \\
4027 & 2455160.9122478 & 2.0 & 10.0 & 2009-11-25.411 & 3 \\
4028 & 2455160.9936251 & 2.0 & 10.0 & 2009-11-25.492 & 3 \\
4039 & 2455161.888785 & 2.0 & 10.0 & 2009-11-26.387 & 3 \\
4040 & 2455161.9701474 & 2.2 & 10.0 & 2009-11-26.469 & 3 \\
4051 & 2455162.8652907 & 2.0 & 10.0 & 2009-11-27.364 & 3 \\
4052 & 2455162.9466637 & 2.2 & 10.0 & 2009-11-27.446 & 3 \\
4629 & 2455209.9010846 & 2.2 & 10.0 & 2010-01-13.400 & 3 \\
8365 & 2455513.9248341 & 2.0 & 10.0 & 2010-11-13.424 & 3 \\
8366 & 2455514.0062088 & 2.0 & 10.0 & 2010-11-13.505 & 3 \\
9578 & 2455612.6349366 & 2.0 & 10.0 & 2011-02-20.134 & 3 \\
9579 & 2455612.7163044 & 2.0 & 10.0 & 2011-02-20.215 & 3 \\
12616 & 2455859.8576278 & 2.0 & 10.0 & 2011-10-25.356 & 3 \\
12617 & 2455859.9390212 & 2.0 & 10.0 & 2011-10-25.438 & 3 \\
23499 & 2456745.48149 & 1.1 & 1.0 & 2014-03-28.981 & 7 \\
23500 & 2456745.56289 & 1.1 & 1.0 & 2014-03-29.062 & 7 \\
23510 & 2456746.37666 & 1.1 & 1.0 & 2014-03-29.876 & 7 \\
23511 & 2456746.45803 & 1.1 & 1.0 & 2014-03-29.957 & 7 \\
23522 & 2456747.35317 & 1.1 & 1.0 & 2014-03-30.852 & 7 \\
23599 & 2456753.6192 & 1.6 & 1.0 & 2014-04-6.118 & 7 \\
23600 & 2456753.70053 & 1.1 & 1.0 & 2014-04-6.199 & 7 \\
23636 & 2456756.63019 & 1.1 & 1.0 & 2014-04-9.129 & 7 \\
23707 & 2456762.40788 & 1.1 & 1.0 & 2014-04-14.907 & 7 \\
25336 & 2456894.9706906 & 2.0 & 10.0 & 2014-08-25.469 & 3 \\
25373 & 2456897.9816518 & 2.0 & 10.0 & 2014-08-28.481 & 3 \\
25385 & 2456898.9581795 & 2.0 & 10.0 & 2014-08-29.457 & 3 \\
25434 & 2456902.9456439 & 2.0 & 10.0 & 2014-09-2.444 & 3 \\
25532 & 2456910.9205721 & 2.0 & 10.0 & 2014-09-10.419 & 3 \\
25545 & 2456911.9784502 & 2.0 & 10.0 & 2014-09-11.477 & 3 \\
25789 & 2456931.8343725 & 2.0 & 10.0 & 2014-10-1.333 & 3 \\
\multicolumn{6}{c}{Available in its entirety in the machine-readable format.}
\enddata
\tablenotetext{a}{Epoch calculated based on the Schwope 2015 ephemeris.}
\tablenotetext{b}{SLKT = 2; McDonald = 4; MONET/N = 1; Calar Alto = 3; LCO = 5; AIP = 6, Schwope et al. (2015) = 7, TESS = 8}
\end{deluxetable}

\end{document}